\documentclass[conference]{IEEEtran}

\usepackage[T1]{fontenc}

\usepackage{amssymb}	 
\usepackage{amsmath}

\usepackage{balance}
\usepackage{graphicx}
\usepackage{booktabs}
\usepackage{multirow}
\usepackage{tabularx}
\usepackage{threeparttable}
\usepackage{url}

\binoppenalty=\maxdimen
\relpenalty=\maxdimen

\begin{document}

\title{Classifying Web Exploits with Topic Modeling}
\author{\IEEEauthorblockN{Jukka Ruohonen}
\IEEEauthorblockA{University of Turku, Finland \\ \texttt{juanruo@utu.fi}}}

\maketitle

\begin{abstract}
This short empirical paper investigates how well topic modeling and database meta-data characteristics can classify web and other proof-of-concept (PoC) exploits for publicly disclosed software vulnerabilities. By using a dataset comprised of over 36 thousand PoC exploits, near a 0.9 accuracy rate is obtained in the empirical experiment. Text mining and topic modeling are a significant boost factor behind this classification performance. In addition to these empirical results, the paper contributes to the research tradition of enhancing software vulnerability information with text mining, providing also a few scholarly observations about the potential for semi-automatic classification of exploits in the existing tracking infrastructures.
\end{abstract}


\section{Introduction}

This short empirical paper investigates whether and how well topic modeling and database meta-data characteristics can classify web and other proof-of-concept (PoC) exploits for publicly disclosed software vulnerabilities. To provide a tentative definition, PoC exploits are software implementations for concretely demonstrating the existence of a software vulnerability, which is a software defect with security implications. While underlining that PoC exploits may perform poorly for actual exploitation \cite{Holm17}, this definition also aligns and connects the scholarly vulnerability research tradition to the field of software repository mining \cite{Buse12, NguyenMassacci15}. However, the existing research in this tradition has mostly focused on software vulnerabilities. This gap in the literature may yield biases because not all vulnerabilities are exploitable as such; hence, not all vulnerabilities are relevant for repository mining. 

The research on automatic solutions for defect labeling and related bug ``triaging'' aspects (e.g., \cite{Anvik11, Zimmermann12}) provide a further motivation. Exploits are disclosed and published on numerous different Internet platforms---including but not limited to security databases, mailing lists, blogs, and bug trackers. Although systematic archiving has recently gained traction, a human is still required to evaluate and classify exploits gathered from such diverse sources. This human-side has largely been also ignored in previous research. For instance, many proposals for unifying vulnerabilities, exploits, and related security information with different abstract frameworks (see \text{\cite{ChengRoschke09, Moreira08, Settani16, Tian03}}) do not discuss how the unification should work in practice. In other words, proposing a yet another ontology or a further database schema is arguably unfruitful when there are well-known problems for managing even the current volume of vulnerabilities and exploits (cf.~\cite{Christey13, OSVDB16}). In addition to benefits for manual exploit triaging, automation improvements are beneficial for large-scale harvesting of exploits with web crawling and related data collection techniques.

The noteworthy related works are the text mining applications for examining and enriching software vulnerability and related software security information \text{\cite{Chen10, Neuhaus10, Scarabeo15, Toloudis16}}. Given this background, the paper examines the natural language characteristics of the exploits archived to the open data Exploit Database (henceforth, EDB) \cite{EDB17a}, which is currently the likely most comprehensive database for archiving exploits. According to surveys, it is also preferred by many contemporary exploit developers \cite{FangHafiz14}. For examining the database and the exploits archived, a small classification experiment is conducted for evaluating how well existing EDB meta-data characteristics and text mining can classify exploits. For brevity, exploits for web application vulnerabilities are classified in relation to other software types targeted by exploits. This empirical classification experiment is prepared in Section~\ref{section: materials and methods}. Results and discussion follow in Sections \ref{section: results}~ and~\ref{section: discussion}.

\section{Materials and Methods}\label{section: materials and methods}

The raw datasets contains all exploits that were archived to EDB in early October 2016. While (mis)selection of a corpora is a typical reason for construct validity biases in topic modeling applications \cite{Marciniak16}, the analysis in this paper exploits a ``natural bias'' in terms of the persistence of different web vulnerabilities \cite{Neuhaus10, Homaei17}, and, hence, web exploits. That is to say, the intention is to classify web and other exploits based on meta-data and natural language characteristics. The meta-data characteristics refer to the database schema used in EDB, which is maintained semi-manually for categorization and other purposes, whereas the latter are derived  by text mining techniques for which pre-processing is also required. 

\subsection{Pre-processing}

The PoC exploits archived to EDB are in raw text format. In many cases, the exploits refer to self-contained programming code snippets that demonstrate the proof-of-concepts for exploitation. The programming languages used include C, Python, Perl, Ruby, and other scripting languages. Sparse code comments are also typically present. However, the issue of separating exploit code from natural language characteristics is not comparable to the well-known (e.g., \cite{Hindle12}) problem of separating code from code comments. Instead, the issue resembles more the genre of software artifact retrieval from heterogeneous semi-structured sources \text{\cite{Ponzanelli15, WuDu14}}. In other words, the exploits archived refer to free-form text entries that contain information beyond programming details. For instance, attribution credits, disclosure details, and remediation instructions may be included in the archives, while the actual exploitation code may amount only to a few lines of code. 



For these reasons, each archived entry is pre-processed without explicitly attempting to separate the non-code text excerpts from the programming language code. Implicitly, however, a partial separation is done during the six pre-processing steps subsequently enumerated.

\begin{enumerate}
\item{Due to the aforementioned issue related to separating exploit code from non-code, it is not reasonable to consider tokenization techniques such as text splitting according to the so-called ``CamelCase'' or ``under\_score'' notations \text{\cite{ChenHassan16, Khatiwada16, Pawelka15}}, or according to the analogous ``slash/notation''~\cite{Maggi11} and ``dot.notation'' \cite{Perdisci12} commonly used for malware labeling, for instance. Thus, each raw exploit entry is first tokenized simply by using the \texttt{word\_tokenize} function in the \textit{de~facto} text ming library for Python~\cite{NLTK}. These tokens are used for inputs to the subsequent pre-processing checks and routines.}
\item{Tokens less than four characters in length are subsequently excluded alongside with tokens with length longer than 20 characters. While the former exclusion criterion is commonly used in text mining (e.g., \cite{Pawelka15}), the latter is specific to the context, typically capturing and excluding tokens referring to large hexadecimal payloads used in exploits for buffer overflow vulnerabilities.}
\item{The next step involves classifying all tokens into words and non-words (henceforth, \textit{words} and \textit{terms}). Python bindings~\cite{pyenchant} for a common open source English dictionary~\cite{enchant} are used for this classification of the tokens.}
\item{The WordNet-based NLTK function \texttt{lemmatize} is then used for grouping similar words (but not terms) according to their dictionary forms. Due to the specific context of exploits, lemmatization is preferable to conventional stemming algorithms. For instance, exemplifying typical over-stemming issues \cite{Khatiwada16}, the word \texttt{vulnerability} is undesirably stemmed to \texttt{vulner}.}
\item{A few stopwords are removed from the collections of lemmatized words (and non-lemmatized terms as a double-check). NLTK's default stopword list is used.}
\item{After all exploits have been processed, those words and terms are excluded that have a frequency less than 20 across all exploits observed. While this again commonly used (e.g., \cite{TehBlei06}) frequency-based pre-processing criterion excludes a substantial amount of words and terms, the choice is partially justified by practical aspects related to computational memory requirements. Finally, seven exploits (alongside the corresponding words and terms) were excluded from the analysis because either no words or no terms were present after the exclusion.}
\end{enumerate}



These six steps result two frequency matrices for words and terms; each row denotes an exploit and each column an unigram (a word or a term), such that the $(i, j)$:th entry contains the frequency of the $j$:th unigram for the $i$:th observed exploit among the $n = 36,184$ exploits observed. In total, $4,844$ words and $7,995$ terms were identified from these exploits with the sixfold pre-processing routine. Although the separation is only implicit, the two matrices can be reflected against the lower entropy that programming code typically exhibits compared to natural language \cite{Hindle12}. The same likely applies also to general technical terms and security industry slang typically used in the sample. Therefore, separate topics are extracted from the two unigram matrices, and the topics computed are used as separate covariates for classification.

\subsection{Topics}\label{subsec: topics}

The topics are extracted from the two frequency matrices by using the latent Dirichlet allocation (LDA) method developed by Blei and associates~\cite{Blei12}. As this method is a modern classic in computer science, which is accompanied by surveys and \text{hands-on} guides for software engineering~\cite{ChenHassan16} and related fields~\cite{Debortoli16}---including detailed information about the use for software vulnerabilities~\cite{Neuhaus10}---a few practical remarks are more relevant than a brief summary of the LDA method itself.

In a nutshell, the method is based on Bayesian mixture modeling in which a topic is a multinomial probability distribution over words from a finite vocabulary \cite{TehBlei06}. As Dirichlet processes are used for deriving the prior distributions in LDA, the first practical concern for applied research relates to the two parameters governing the per-exploit topic distribution and per-topic ``word-or-term'' distribution. Although these parameters should arguably reflect genuine prior information, computational routines and rules-of-thumb are commonly used in practice~\text{\cite{ChenHassan16, Debortoli16}}. In this paper, likewise, the estimation routine in the R package used for computation~\cite{topicmodels} is adopted for determining the parameters (i.e., the package's default settings are used). The second issue for applied work relates to the fact that a single exploit is often characterized by multiple dominant topics, which entails a choice over a threshold scalar for cutting off irrelevant topics~\cite{BaruaHassan14}. Given the classification purposes of this paper, each exploit is assigned to the most dominant topic with the highest membership rate. The third and final concern relates to the number of topics to extract. Because the topics assigned for each exploit are used for classification, which implies that interpretation and construct validity are lesser concerns \cite{Debortoli16}, the LDA is computed, for the term and frequency matrices separately, by restricting the number of topics to $k = 5, 10, 20, 30, 40, 50$. For each of these restrictions, the corresponding dominant (word and term) topic assignments vectors are used to build separate classifiers.

\subsection{Classification}

The classification experiment is computed by using an R implementation \cite{randomForest}, in combination with the \textit{caret} package~\cite{caret}, for the tree-based random forest classifier. Rather than summarizing this decision tree method, it is again more relevant to focus on the practical aspects, starting from the operationalization of the two binary-valued response metrics.

\subsubsection{Responses}\label{subsec: responses}

\begin{figure}[th!b]
\centering
\includegraphics[width=\linewidth, height=2.5cm]{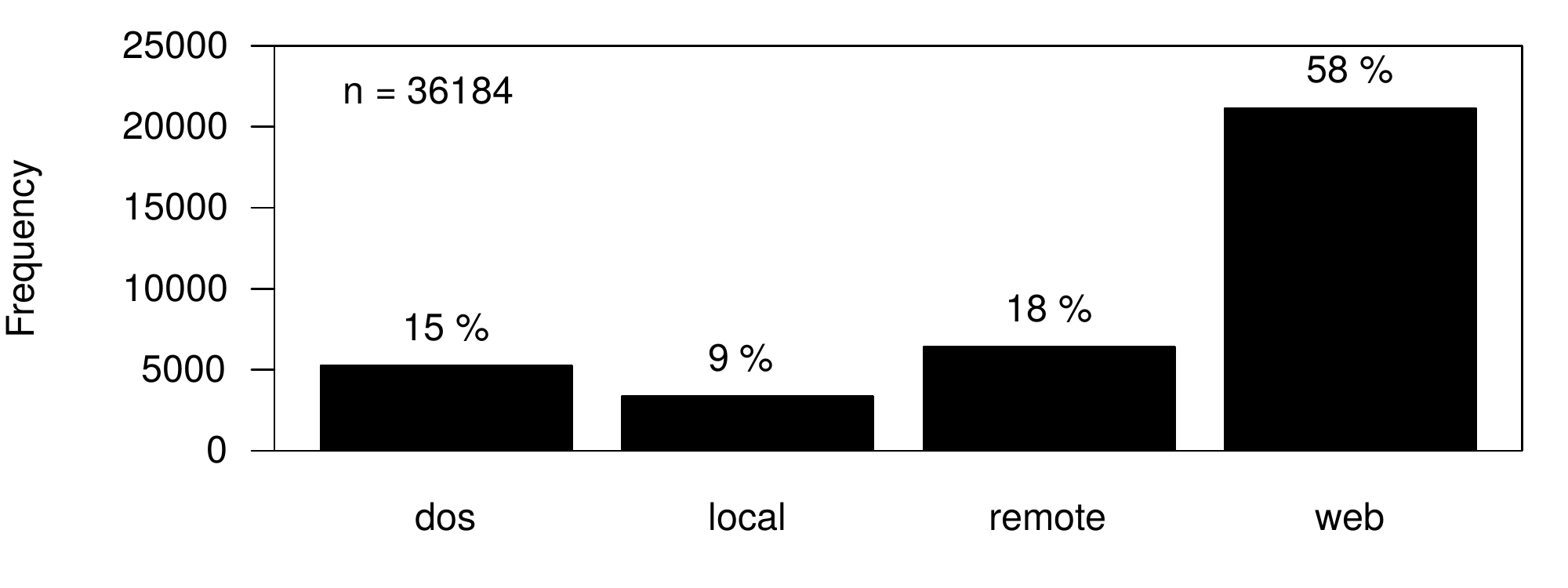}
\caption{Exploit Categories (EDB's meta-data)}
\label{fig: category}
\end{figure}

The binary-valued response metrics for the classification are constructed from two meta-data schemas provided in EDB. The first is based on the high-level categorization into denial-of-service (dos), local, remote, and web exploits. As can be observed from Fig.~\ref{fig: category}, about 58 percent of all exploits observed are located in the web-category. For the first response metric, these web exploits are used as a reference category (``true''), while the remaining three categories are grouped into a single group of non-web exploits (``false'').

\begin{figure}[th!b]
\centering
\includegraphics[width=\linewidth, height=3.2cm]{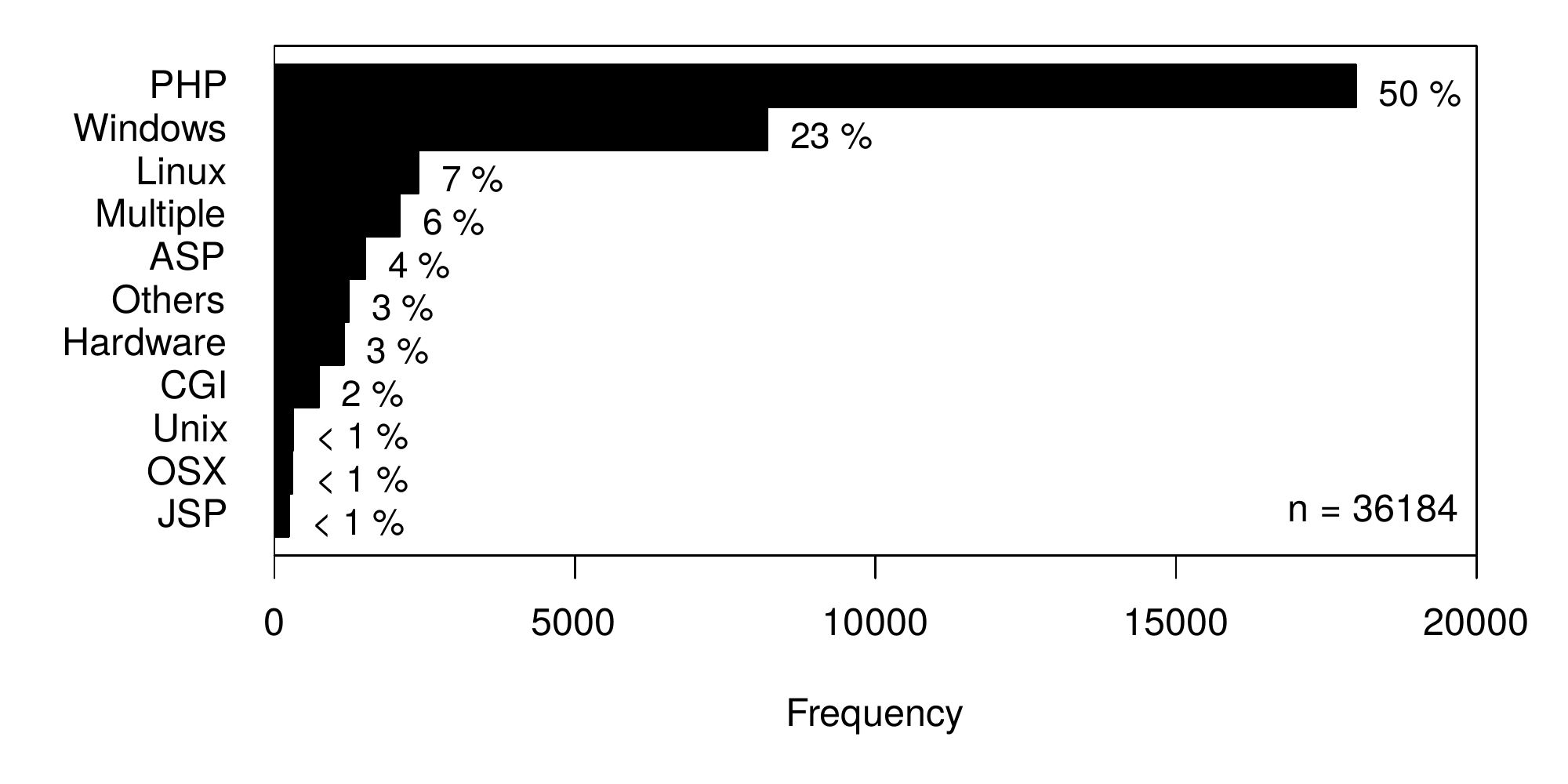}
\caption{Exploit Platforms (EDB's meta-data)}
\label{fig: platform}
\end{figure}

The second response metric is based on the target platforms to which the EDB community additionally groups the exploits archived. In total, as many as 51 distinct platforms have been used for this grouping, although the most common platforms show no big surprises. By regrouping less common platforms into a heterogeneous group of ``Others'', the ten most frequent platforms are illustrated in Fig.~\ref{fig: platform}. Reflecting the large amount of web exploits in the earlier Fig.~\ref{fig: category}, web sites and web applications written in PHP have been the most frequent targets for the exploits archived, followed by software running on Windows and Linux, respectively. Analogous to the first response metric, the PHP platform is taken as a reference category against which the remaining platforms are compared.

\subsubsection{Covariates}

\begin{table*}[t!]
\centering
\caption{Covariates}
\label{tab: covariates}
\begin{tabular}{lll}
\toprule
\# & Mnemonic name & Description \\
\hline
1. & \textit{Term topic} & One for the most dominant term-based topic characterizing the exploit; zero otherwise. \\
\cmidrule{3-3}
2. & \textit{Word topic} & One for the most dominant word-based topic characterizing the exploit; zero otherwise. \\
\cmidrule{3-3}
3. & \textit{Verified} & One if the EDB community has verified the exploit; zero otherwise. \\
\cmidrule{3-3}
4. & \textit{Application} & One if the vulnerable application is available for download; zero otherwise. \\
\cmidrule{3-3}
5. & \textit{Screenshot} & One if a screenshot is provided for a demonstration or other purposes; zero otherwise. \\
\cmidrule{3-3}
6. & \textit{OSVDB references} & The number of OSVDB references or zero for no such references. \\
\cmidrule{3-3}
7. & \textit{CVE references} & The number of CVE references or zero for the absence of CVE references. \\
\cmidrule{3-3}
8. & \textit{Mean CVSS} & The mean of CVSS base scores for all CVE references (or zero for no references).\\
\cmidrule{3-3}
9. & \textit{Publication year} & The year at which the exploit was first published according to EDB. \\
\cmidrule{3-3}
10. & \textit{Publication month} & The month at which the exploit was first published according to EDB. \\
\cmidrule{3-3}
11. -- 40.~& \textit{Top developers} & One if the author of the exploit is among the ``top-30'' developers; zero otherwise. \\
\bottomrule
\end{tabular}
\end{table*}

The covariate metrics enumerated in Table~\ref{tab: covariates} are included for classifying web and PHP exploits. While the first two metrics are based on the dominant topic assignment vectors (see Section~\ref{subsec: topics}), the remaining ones are all based on EDB's meta-data. These meta-data covariates can be categorized into four high-level abstract groups.

\begin{enumerate}
\item{The first group includes three binary-valued metrics related to the evaluation work done by the EDB community for the exploits archived. Like all software, exploits require testing. \textit{Verified}, therefore, scores one in case the exploit has been verified to work. \textit{Application} is likewise a dummy variable for recording whether the archive contains also the vulnerable application available for download. Reflecting the occasional use of video demonstrations during software vulnerability disclosure~\cite{Google16}, the third dummy variable records the cases for which a screenshot is available for download.}
\item{The second group contains three metrics explicitly or implicitly related to other databases. Akin to the ``vulnerability claims'' made by the orchestrators of the National Vulnerability Database (NVD) \cite{NguyenMassacci15}, per-exploit references made by the EDB community to the Open Source Vulnerability Database (OSVDB) are counted with the metric named \textit{OSVDB references}. (It can be also remarked that OSVDB used to be the only comprehensive open database containing some information about exploits \cite{Massacci10}, but this database was shutdown in 2016 due to maintenance and other issues \cite{OSVDB16}. This does not affect the results reported, however.) For each exploit, the amount of references to Common Vulnerabilities and Exposures (CVEs) are likewise counted via the metric labeled \textit{CVE references}. While these reference counters are implicit in the sense that no attempts are made to validate EDB's claims, the Common Vulnerability Scoring System (CVSS) is explicitly used to approximate the average severity of the vulnerabilities targeted by those exploits that have CVE references. \textit{Mean CVSS} thus denotes the arithmetic mean of so-called base CVSS scores, which range from zero (negligible) to ten (catastrophic), and which are explicitly fetched from~NVD.}
\item{The third group contains two calendar time metrics, \textit{Month} and \textit{Year}. Due to the well-known data quality issues affecting vulnerability (e.g., \cite{Christey13, Massacci10}) and exploit archives, interpretation should be done with care--- but, nevertheless, these two metrics convey the approximate year and month at which a given exploit was first published in the wild according to EDB's evaluation.}
\item{The final set of covariates expands to 30 dummy variables for the most productive exploit developers. Thus, the first dummy variable scores one for all exploits authored by the most productive developer, and so forth. When ranked by the number of exploits authored, the resulting list largely confirms existing empirical observations \text{\cite{Christey13, Johnson16, Ruohonen16RCIS}}. For instance, the Meta\-sploit project and Luigi Auriemma are the top-two most productive authors, but the list contains also names such as rgod, r0t, and the Google Security Research group.}
\end{enumerate}

\section{Results}\label{section: results}

Estimation is carried by fitting $2 \times 7$ models for classifying web and PHP exploits (see Section~\ref{subsec: responses}). In terms of practical estimation concerns, PHP exploits yield a precisely balanced set (see Fig.~\ref{fig: platform}). A sufficiently balanced set is available also for web exploits (see Fig.~\ref{fig: category}). For comparing how much information is gained from the dominant topic assignments (see Section~\ref{subsec: topics}), classification of the two response metrics is first done by excluding the topic assignment vectors, and then fitting separate models for the varying number of topics extracted via LDA. As the topic assignments vectors for words and terms do not notably correlate with each other (see Table~\ref{tab: correlations}), both vectors are included in these separate models. A $5$-fold cross-validation is used for training. To maintain a sensible scenario for predicting new data, the exploits published in 2016 are used as a test set; these amount to about 3.5~\% of all exploits observed. (It can be remarked that comparable results are obtained with randomly picked test sets containing 10~\% of the exploits.) Finally, accuracy (i.e., the number of true positives and true negatives to all exploits observed) is sufficient as a performance evaluation metric in this paper. The accuracy rates reported in Table~\ref{tab: classification} are presented also with 95~\% confidence intervals (CIs).

\begin{table}[t!]
\centering
\caption{Topic Assignment Correlations (words and terms)}
\label{tab: correlations}
\begin{tabular}{lcccccc}
\toprule
& \multicolumn{6}{c}{Topics ($k$)} \\
\cmidrule{2-7}
& 5 & 10 & 20 & 30 & 40 & 50 \\
\cmidrule{2-7}
Spearman rho & $-0.16$ & $0.23$ & $0.08$ & $0.22$ & $0.06$ & $-0.10$ \\
\bottomrule
\end{tabular}
\end{table}

\begin{table}[t!]
\centering
\caption{Classification Performance}
\label{tab: classification}
\begin{tabular}{lcccc}
\toprule
\multirow{3}{*}{$k$} &
\multirow{3}{*}{Covariates} & \multicolumn{3}{c}{Accuracy} \\
\cmidrule{3-5}
&& Web~\scriptsize{[95~\% CIs]} && PHP~\scriptsize{[95~\% CIs]} \\
\hline
\phantom{1}0 & 38 & 0.788~\scriptsize{[0.765, 0.810]} && 0.742~\scriptsize{[0.717, 0.766]} \\
\phantom{1}5 & 40 & 0.895~\scriptsize{[0.877, 0.911]} && 0.843~\scriptsize{[0.821, 0.862]} \\
10 & 40 & 0.910~\scriptsize{[0.893, 0.925]} && 0.861~\scriptsize{[0.841, 0.880]} \\
20 & 40 & 0.920~\scriptsize{[0.904, 0.935]} && 0.888~\scriptsize{[0.869, 0.905]} \\
30 & 40 & 0.912~\scriptsize{[0.894, 0.927]} && 0.881~\scriptsize{[0.862, 0.898]} \\
40 & 40 & 0.914~\scriptsize{[0.897, 0.929]} && 0.863~\scriptsize{[0.843, 0.882]} \\
50 & 40 & 0.913~\scriptsize{[0.896, 0.928]} && 0.878~\scriptsize{[0.858, 0.895]} \\
\bottomrule
\end{tabular}
\end{table}

According to the results, performance is rather good for both response metrics, although higher accuracy rates are obtained for web exploits. Performance also increases from the inclusion of the dominant topic vectors. Given the optimum number of topics around $k \simeq 20$, the performance increases are about $0.132$ and $0.146$ for web and PHP exploits, respectively.

\section{Discussion}\label{section: discussion}

This short empirical paper examined how well database meta-data and topic modeling can classify web-related and other exploits for known software vulnerabilities. By using a dataset of over 36 thousand exploits, LDA with varying number of topics, and a random forest classifier, the overall accuracy rate was observed to be in the range $[0.89, 0.92]$, which is a fairly decent range in empirical software engineering applications. The few remaining points can be presented in terms of limitations, which also provide more focused questions for further empirical research.

While topic modeling was shown to provide a neat dimensional reduction technique for classification, (a) it is unclear whether better performance might have been obtained by using the raw frequency matrices (or the inverse variants). In the same vein, (b) the classification experiment was limited to web and PHP exploits, although practical real-world applications would likely require multi-class classification that takes all categories (cf.~Fig.~\ref{fig: category}) and platforms (cf.~Fig.~\ref{fig: platform}) into account.

Moreover, (c) different thresholds \cite{ChenHassan16} could be adopted for the dominant topic assignments. Perhaps a more important concern relates to the pre-processing routines, which---like arguably in most text mining applications---are always exposed to validity concerns. In particular, (d) the used separation between English words and non-words is only a coarse approximation insofar as actual exploitation programming code is considered. In this respect, exploits posit a particularly challenging case for mining and extracting software engineering artifacts from heterogeneous sources \cite{Ponzanelli15, WuDu14, Pawelka15}. Such extraction would have also practical value. Finally, (e) the near 0.9 accuracy rate can be still argued to be modest because most of the predictive power still comes from the meta-data characteristics, which require manual, human-made classification and related evaluation work. Taken together, these observations require further research on automatic classification of vulnerabilities and exploits gathered from heterogeneous Internet sources.

\section*{Acknowledgments}

The author acknowledges Tekes---the Finnish Funding Agency for Innovation, DIMECC Oy, and the Cyber Trust research program for their support. 

\balance
\bibliographystyle{IEEEtran}

\begin{thebibliography}{10}
\providecommand{\url}[1]{#1}
\csname url@samestyle\endcsname
\providecommand{\newblock}{\relax}
\providecommand{\bibinfo}[2]{#2}
\providecommand{\BIBentrySTDinterwordspacing}{\spaceskip=0pt\relax}
\providecommand{\BIBentryALTinterwordstretchfactor}{4}
\providecommand{\BIBentryALTinterwordspacing}{\spaceskip=\fontdimen2\font plus
\BIBentryALTinterwordstretchfactor\fontdimen3\font minus
  \fontdimen4\font\relax}
\providecommand{\BIBforeignlanguage}[2]{{%
\expandafter\ifx\csname l@#1\endcsname\relax
\typeout{** WARNING: IEEEtran.bst: No hyphenation pattern has been}%
\typeout{** loaded for the language `#1'. Using the pattern for}%
\typeout{** the default language instead.}%
\else
\language=\csname l@#1\endcsname
\fi
#2}}
\providecommand{\BIBdecl}{\relax}
\BIBdecl

\bibitem{Holm17}
H.~Holm and T.~Sommestad, ``{S}o {L}ong, and {T}hanks for {O}nly {U}sing
  {R}eadily {A}vailable {S}cripts,'' \emph{Information \& Computer Security},
  vol.~25, no.~1, pp. 47--61, 2017.

\bibitem{Buse12}
R.~P.~L. Buse and T.~Zimmermann, ``{I}nformation {N}eeds for {S}oftware
  {D}evelopment {A}nalytics,'' in \emph{Proceedings of the 34th International
  Conference on Software Engineering (ICSE 2012)}.\hskip 1em plus 0.5em minus
  0.4em\relax Zurich: IEEE, 2012, pp. 987--996.

\bibitem{NguyenMassacci15}
V.~H. Nguyen, S.~Dashevskyi, and F.~Massacci, ``{A}n {A}utomatic {M}ethod for
  {A}ssessing the {V}ersions {A}ffected by a {V}ulnerability,'' \emph{Empirical
  Software Engineering}, vol.~21, no.~6, pp. 2268--2297, 2015.

\bibitem{Anvik11}
J.~Anvik and G.~C. Murphy, ``{R}educing the {E}ffort of {B}ug {R}eport
  {T}riange: {R}ecommenders for {D}evelopment-{O}riented {D}ecisions,''
  \emph{ACM Transactions on Software Engineering and Methodology}, vol.~20,
  no.~3, pp. 10:1--10:35, 2011.

\bibitem{Zimmermann12}
T.~Zimmermann, N.~Nagappan, P.~J. Guo, and B.~Murphy, ``{C}haracterizing and
  {P}redicting {W}hich {B}ugs {G}et {R}eopened,'' in \emph{Proceedings of the
  34th International Conference on Software Engineering (ICSE 2012)}.\hskip 1em
  plus 0.5em minus 0.4em\relax Piscataway: IEEE, 2012, pp. 1074--1083.

\bibitem{ChengRoschke09}
F.~Cheng, S.~Roschke, R.~Schuppenies, and C.~Meinel, ``{R}emodeling
  {V}ulnerability {I}nformation,'' in \emph{Proceedings of the International
  Conference on Information Security and Cryptology (Inscrypt 2009}, F.~Bao,
  M.~Yung, D.~Lin, and J.~Jing, Eds.\hskip 1em plus 0.5em minus 0.4em\relax
  Beijing: Springer, 2009, pp. \text{324--336}.

\bibitem{Moreira08}
E.~dos Santos~Moreira, L.~A.~F. Martimiano, A.~J. dos Santos Brand\~ao, and
  M.~C. Bernardes, ``{O}ntologies for {I}nformation {S}ecurity {M}anagement and
  {G}overnance,'' \emph{Information Management \& Computer Security}, vol.~16,
  no.~2, pp. 105--165, 2008.

\bibitem{Settani16}
G.~Settanni, F.~Skopik, Y.~Shovgenya, R.~Fiedler, M.~Carolan, D.~Conroy,
  K.~Boettinger, M.~Gall, G.~Brost, C.~Ponchel, M.~Haustein, H.~Kaufmann,
  K.~Theuerkauf, and P.~Olli, ``{A} {C}ollaborative {C}yber {I}ncident
  {M}anagement {S}ystem for {E}uropean {I}nterconnected {C}ritical
  {I}nfrastructures,'' \emph{Journal of Information Security and Applications},
  no. Published online in 2 June 2016, 2016.

\bibitem{Tian03}
H.~Tian, L.~Huang, Z.~Zhou, and H.~Zhang, ``{C}ommon {V}ulnerability {M}arkup
  {L}anguage,'' in \emph{{P}roceedings of the {F}irst {I}nternational
  {C}onference on {A}pplied {C}ryptography and {N}etwork {S}ecurity (ACNS
  2003), {L}ecture {N}otes in {C}omputer {S}cience (Volume 2846)}, J.~Zhou,
  M.~Yung, and Y.~Han, Eds.\hskip 1em plus 0.5em minus 0.4em\relax Kunming:
  Springer, 2003, pp. 228--240.

\bibitem{Christey13}
S.~Christey and B.~Martin, ``{B}uying {I}nto the {B}ias: {W}hy {V}ulnerability
  {S}tatistics {S}uck,'' in \emph{Presentation at Black Hat 2013}, Las Vegas,
  2013, available online in January 2017:
  \url{https://media.blackhat.com/us-13/US-13-Martin-Buying-Into-The-Bias-Why-Vulnerability-Statistics-\\Suck-Slides.pdf}.

\bibitem{OSVDB16}
{Open Source Vulnerability Database}, ``{OSVDB: FIN},'' 2016, available online
  in February 2017: \url{https://blog.osvdb.org/2016/04/05/osvdb-fin/}.

\bibitem{Chen10}
Z.~Chen, Y.~Zhang, and Z.~Chen, ``{A} {C}ategorization {F}ramework for {C}ommon
  {C}omputer {V}ulnerabilities and {E}xposures,'' \emph{The Computer Journal},
  vol.~53, no.~5, pp. \text{551--580}, 2010.

\bibitem{Neuhaus10}
S.~Neuhaus and T.~Zimmermann, ``{S}ecurity {T}rend {A}nalysis with {CVE}
  {T}opic {M}odels,'' in \emph{Proceedings of the IEEE 21st International
  Symposium on Software Reliability Engineering (ISSRE 2010)}.\hskip 1em plus
  0.5em minus 0.4em\relax San Jose: IEEE, 2010, pp. 111--120.

\bibitem{Scarabeo15}
N.~Scarabeo, B.~C. Fung, and R.~H. Khokhar, ``{M}ining {K}nown {A}ttack
  {P}atterns from {S}ecurity-{R}elated {E}vents,'' \emph{PeerJ Computer
  Science}, vol. e25, no.~1, pp. \text{1--21}, 2014.

\bibitem{Toloudis16}
D.~Toloudis, G.~Spanos, and L.~Angelis, ``{A}ssociating the {S}everity of
  {V}ulnerabilities with {T}heir {D}escription,'' in \emph{Proceedings of the
  CAiSE 2016 International Workshops, Lecture Notes in Business Information
  Processing (Volume 249)}, J.~Krogstie, H.~Mouratidis, and J.~S.~D. Toloudis,
  Eds.\hskip 1em plus 0.5em minus 0.4em\relax Ljubljana: Springer, 2016, pp.
  \text{231--242}.

\bibitem{EDB17a}
{Exploit Database}, ``{O}ffensive {S}ecurity {E}xploit {D}atabase,'' 2017,
  available online in February 2017: \url{https://www.exploit-db.com/}.

\bibitem{FangHafiz14}
M.~Fang and M.~Hafiz, ``{D}iscovering {B}uffer {O}verflow {V}ulnerabilities in
  the {W}ild: {A}n {E}mpirical {S}tudy,'' in \emph{Proceedings of the 8th
  ACM/IEEE International Symposium on Empirical Software Engineering and
  Measurement (ESEM 2014)}.\hskip 1em plus 0.5em minus 0.4em\relax Torino: ACM,
  2014, pp. 1--10.

\bibitem{Marciniak16}
D.~Marciniak, ``{C}omputational {T}ext {A}nalysis: {T}houghts on the
  {C}ontingencies of an {E}volving {M}ethod,'' \emph{Big Data \& Society}, pp.
  1--5, 2016.

\bibitem{Homaei17}
H.~Homaei and H.~R. Shahriari, ``{S}even {Y}ears of {S}oftware
  {V}ulnerabilities: {T}he {E}bb and {F}low,'' \emph{Security \& Privacy},
  vol.~15, no.~1, pp. 58--65, 2017.

\bibitem{Hindle12}
A.~Hindle, E.~T. Barr, Z.~Su, M.~Gabel, and P.~Devanbu, ``{O}n the
  {N}aturalness of {S}oftware,'' in \emph{Proceedings of the 34th International
  Conference on Software Engineering (ICSE 2012)}.\hskip 1em plus 0.5em minus
  0.4em\relax Zurich: IEEE, 2012, pp. 837--847.

\bibitem{Ponzanelli15}
L.~Ponzanelli, A.~Mocci, and M.~Lanza, ``{S}ummarizing {C}omplex {D}evelopment
  {A}rtifacts by {M}ining {H}eterogeneous {D}ata,'' in \emph{Proceedings of the
  12th Working Conference on Mining Software Repositories (MSR 2015)}.\hskip
  1em plus 0.5em minus 0.4em\relax Piscataway: IEEE, 2015, pp. 401--405.

\bibitem{WuDu14}
L.~Wu, L.~Du, B.~Liu, G.~Xu, Y.~Ge, Y.~Fu, J.~Li, Y.~Zhou, and H.~Xiong,
  ``{H}eterogeneous {M}etric {L}earning with {C}ontent-{B}ased {R}egularization
  for {S}oftware {A}rtifact {R}etrieval,'' in \emph{Proceedings of the IEEE
  International Conference on Data Mining (ICDM 2014)}.\hskip 1em plus 0.5em
  minus 0.4em\relax Shenzhen: IEEE, 2014, pp. 610--619.

\bibitem{ChenHassan16}
T.-H. Chen, S.~W. Thomas, and A.~E. Hassan, ``{A} {S}urvey on the {U}se of
  {T}opic {M}odels {W}hen {M}ining {S}oftware {R}epositories,'' \emph{Empirical
  Software Engineering}, vol.~21, no.~5, pp. 1843--1919, 2016.

\bibitem{Khatiwada16}
S.~Khatiwada, M.~Kelly, and A.~Mahmoud, ``{STAC}: {A} {T}ool for {S}tatic
  {T}extual {A}nalysis of {C}ode,'' in \emph{Proceedings of the IEEE 24th
  International Conference on Program Comprehension (ICPC 2016)}.\hskip 1em
  plus 0.5em minus 0.4em\relax Austin: IEEE, 2016, pp. 1--3.

\bibitem{Pawelka15}
T.~Pawelka and E.~Juergens, ``{I}s {T}his {C}ode {W}ritten in {E}nglish? {A}
  {S}tudy of the {N}atural {L}anguage of {C}omments and {I}dentifiers in
  {P}ractice,'' in \emph{Proceedings of the IEEE International Conference on
  Software Maintenance and Evolution (ICSME 2015)}.\hskip 1em plus 0.5em minus
  0.4em\relax Bremen: IEEE, 2015, pp. 401--410.

\bibitem{Maggi11}
F.~Maggi, A.~Bellini, G.~Salvaneschi, and S.~Zanero, ``{F}inding
  {N}on-{T}rivial {M}alware {N}aming {I}nconsistencies,'' in \emph{Proceedings
  of the 7th International Conference (ICISS 2011), Lecture Notes in Computer
  Science (Volume 7093)}, S.~Jajodia and C.~Mazumdar, Eds.\hskip 1em plus 0.5em
  minus 0.4em\relax Kolkata: Springer, 2011, pp. \text{144--159}.

\bibitem{Perdisci12}
R.~Perdisci and M.~U, ``{VAMO}: {T}owards a {F}ully {A}utomated {M}alware
  {C}lustering {V}alidity {A}nalysis,'' in \emph{Proceedings of the 28th Annual
  Computer Security Applications Conference (ACSAC 2012)}.\hskip 1em plus 0.5em
  minus 0.4em\relax Orlando: ACM, 2012, pp. \text{329--338}.

\bibitem{NLTK}
{T}he {N}atural {L}anguage~{T}oolkit (NLTK), ``{NLTK} 3.0 {D}ocumentation,''
  2017, available online in January 2017: \url{http://www.nltk.org}.

\bibitem{pyenchant}
R.~Kelly, ``{PyEnchant}: {A} {S}pellchecking {L}ibrary for {P}ython (1.6.8),''
  2017, available online in January 2017:
  \url{http://pythonhosted.org/pyenchant/}.

\bibitem{enchant}
D.~Lachowicz, ``{E}nchant (1.6.0),'' 2017, available online in January 2017:
  \url{http://www.abisource.com/projects/enchant/}.

\bibitem{TehBlei06}
Y.~W. Teh, M.~I. Jordan, M.~J. Beal, and D.~M. Blei, ``{H}ierarchical
  {D}irichlet {P}rocesses,'' \emph{Journal of the American Statistical
  Association}, vol. 101, no. 476, pp. 1566--1581, 2006.

\bibitem{Blei12}
D.~M. Blei, ``{P}robabilistic {T}opic {M}odels,'' \emph{Communications of the
  ACM}, vol.~55, no.~4, pp. 77--84, 2012.

\bibitem{Debortoli16}
S.~Debortoli, O.~M\"uller, I.~Junglas, and J.~{vom Brocke}, ``{T}ext {M}ining
  for {I}nformation {S}ystems {R}esearchers: {A}n {A}nnotated {T}opic
  {M}odeling {T}utorial,'' \emph{Communications of the Association for
  Information Systems}, vol.~39, p. Article~7, 2016.

\bibitem{topicmodels}
B.~Gr\"un and K.~Hornik, ``{topicmodel}: {A}n {R} {P}ackage for {F}itting
  {T}opic {M}odels,'' \emph{Journal of Statistical Software}, vol.~40, no.~13,
  pp. 1--30, 2016.

\bibitem{BaruaHassan14}
A.~Barua, S.~W. Thomas, and A.~E. Hassan, ``{W}hat {A}re {D}evelopers {T}alking
  {A}bout? {A}n {A}nalysis of {T}opics and {T}rends in {S}tack {O}verflow,''
  \emph{Empirical Software Engineering}, vol.~19, no.~3, pp. \text{619--654},
  2014.

\bibitem{randomForest}
A.~Liaw and M.~Wiener, ``{C}lassification and {R}egression by {randomForest},''
  \emph{R News}, vol.~2, no.~3, pp. 18--22, 2002.

\bibitem{caret}
M.~Kuhn, ``{B}uilding {P}redictive {M}odels in {R} {U}sing the \texttt{caret}
  {P}ackage,'' \emph{Journal of Statistical Software}, vol.~28, no.~5, pp.
  1--26, 2008.

\bibitem{Google16}
{Google, Inc.}, ``{A}void {V}ideos... {B}ut {I}f {Y}ou {C}an't, {H}ere {A}re
  {S}ome {T}ips {:-)},'' 2016, {G}oogle {B}ughunter {U}niversity, available
  online in November 2016:
  \url{https://sites.google.com/site/bughunteruniversity/improve/how-to-record-an-effective-proof-of-concept-video}.

\bibitem{Massacci10}
F.~Massacci and V.~H. Nguyen, ``{W}hich {I}s the {R}ight {S}ource for
  {V}ulnerability {S}tudies? {A}n {E}mpirical {A}nalysis on {M}ozilla
  {F}irefox,'' in \emph{Proceedings of the 6th International Workshop on
  Security Measurements and Metrics (MetriSec 2010)}.\hskip 1em plus 0.5em
  minus 0.4em\relax Bolzano: ACM, 2010, pp. 4:1--4:8.

\bibitem{Johnson16}
P.~Johnson, D.~Gorton, R.~Langerstr\"om, and M.~Ekstedt, ``{T}ime {B}etween
  {V}ulnerability {D}isclosures: {A} {M}easure of {S}oftware {P}roduct
  {V}ulnerability,'' \emph{Computers \& Security}, vol.~62, pp.
  \text{278--295}, 2016.

\bibitem{Ruohonen16RCIS}
J.~Ruohonen, S.~Hyrynsalmi, and V.~Lepp\"anen, ``{T}rading {E}xploits {O}nline:
  {A} {P}reliminary {C}ase {S}tudy,'' in \emph{Proceedings of the IEEE Tenth
  International Conference on Research Challenges in Information Science (RCIS
  2016)}.\hskip 1em plus 0.5em minus 0.4em\relax Grenoble: IEEE, 2016, pp.
  1--12.

\end{thebibliography}


\end{document}